\documentclass[aps,twocolumn,prl,superscriptaddress,preprintnumbers,showpacs,tightenlines]{revtex4}
\usepackage{amssymb}
\usepackage{amssymb}
\usepackage{epsfig,graphicx,times}
\usepackage{enumerate}
\usepackage{dcolumn}
\usepackage{color}
\usepackage{bm}

\newcommand{\beq}{\begin{equation}}
\newcommand{\eeq}{\end{equation}}

\newcommand{\bea}{\begin{eqnarray}}
\newcommand{\eea}{\end{eqnarray}}

\begin{document}


\title{Quantum heat transfer: A Born Oppenheimer method} 
\author{Lian-Ao Wu}
\affiliation{Department of Theoretical Physics and History of Science, The Basque Country
University (EHU/UPV) and IKERBASQUE -- Basque Foundation for Science, 48011,
Bilbao,Spain}
\author{Dvira Segal}
\affiliation{Chemical Physics Group, Department of Chemistry and Center for Quantum
Information and Quantum Control, University of Toronto, 80 St. George
street, Toronto, Ontario, M5S 3H6, Canada}

\begin{abstract}
We develop a Born-Oppenheimer type formalism for the description of quantum thermal transport 
along hybrid nanoscale objects.
Our formalism is suitable for treating heat transfer in the off-resonant regime,
where e.g., the relevant vibrational modes of the interlocated molecule
are high relative to typical bath frequencies, and at low temperatures
when tunneling effects dominate.
A general expression for the thermal energy current is accomplished, in the form of a generalized Landauer formula.
In the harmonic limit this expression reduces to the standard Landauer result for heat transfer,
while in the presence of nonlinearities multiphonon tunneling effects are realized.
\end{abstract}


\maketitle

\textit{Introduction.---} 
Thermal transport in molecular objects has recently become a 
topic of major  interest, relevant for designing electronic and mechanical nanoscale devices \cite{Pop}, 
and for resolving mechanisms and pathways of energy flow in biomolecules \cite{Leitnerbook}.
In modelling such systems we typically consider an impurity object, a subsystem,
e.g., an alkane molecule  \cite{Segalman}, bridging two thermal reservoirs, 
representing solids or large residues in a protein, maintained each at a fixed temperature.
%
Various treatments have been developed for simulating the
thermal conduction properties of such structures, either classically \cite{Lepri},
or in the quantum regime \cite{Dhar,Tu}.
Among these treatments are the generalized Langevin equation method \cite{Segalcond,Lanj},
the Kinetic-Boltzmann theory \cite{Spohn}, mode coupling theory \cite{MC1},
the non-equilibrium Green's function technique \cite{Tu,Green}, 
classical \cite{Lepri} and mixed classical-quantum \cite{Wang,Tu} molecular dynamics simulations,
and exact quantum simulations on simplified models \cite{Hartree}.

The master equation technique at weak system-bath coupling is of particular interest \cite{MasterD}, 
allowing to obtain simple analytical results in interesting limits \cite{HanggiF}, 
guiding experimentalists and motivating theoreticians in developing more detailed 
treatments \cite{Hartree}. In this approach the heat current is described by sequential {\it incoherent}
emission and absorption processes, relaying on a resonance condition.
Thus, a finite conductance exists only when the frequencies of the
two thermal reservoirs match the subsystem characteristic
frequency.
Nevertheless, in many systems the characteristic frequencies of the impurity object 
are high relative to the cutoff frequencies of the reservoirs. For instance, 
consider an electronic spin surrounded by nuclear spins subjected to an external field, a molecule
of high vibrational frequency coupled to solids with low Debye frequencies, or a high-frequency
heat source inside a protein with low frequency bonds as thermometers \cite{Hamm}.
Developing a general formalism that can treat such scenarios,
providing simple analytical results and bringing in
physical insight, is of a great importance.

Here we describe a new formalism for treating quantum thermal transport in such
{\it non-resonant} systems, where subsystem's 
frequencies, relevant for thermal transfer, are {\it above} the reservoirs spectral window,
or the baths temperatures are low, below the subsystem energy spacing \cite{comment}.
Using a Born-Oppenheimer (BO) type approximation we develop a compact expression
for the heat current in the form of a generalized Landauer formula \cite{Land}. 
For harmonic systems we recover the elastic Landauer formula. When nonlinear interactions persist
multiphonon  processes contribute to the thermal current.

\begin{figure}
\vspace{-14mm}
{\hbox{\epsfxsize=105mm \epsffile{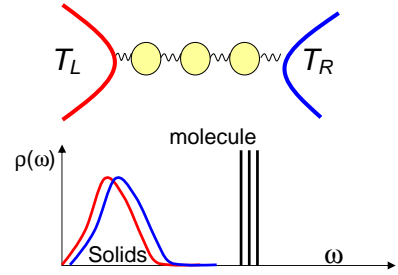}}}
\vspace{-85mm}
\caption{A Scheme of our setup (top), 
including a subsystem, e.g., a molecular chain, connecting two solids.
The bottom panel exemplifies the vibrational spectra $\rho(\omega)$ of the isolated solids and molecule.} 
\label{FigD}
\end{figure}

\textit{Model.--- }
Consider a small subsystem, representing e.g., a molecule, placed
in between two thermal reservoirs (e.g., solids) 
maintained each at a fixed temperatures $T_{\nu}$ ($\nu=L,R$), see Fig. \ref{FigD}.
The total Hamiltonian is given by
\bea
H=H_{S}+H_{L}+H_{R}+V_{L}+V_{R},
\label{eq:H}
\eea
where $H_{S}$ is the Hamiltonian of the subsystem and $H_{\nu}$
stands for the $\nu$ heat bath. $V_{L}$ ($V_{R}$) couples separately 
the subsystem and the left (right)
reservoir. The subsystem and the two reservoirs' Hamiltonians assume
diagonal forms, and we consider a bipartite form,
$V_{\nu}=SB_{\nu}$.
%
Here $S$ is a subsystem operator and $B_{\nu}$ is an operator in
terms of the $\nu$ bath degrees of freedom. In particular, it is useful to study
two extreme realizations for the subsystem. In the first case 
nonlinear interactions are incorporated by adopting a two-level system (TLS)
model \cite{MasterD}, $S=\sigma_{x}$,  $H_{S}=\frac{\epsilon
}{2}\sigma_{z}$. In the second model the subsystem is represented by
a single harmonic mode linearly coupled to the baths, 
$S=b+b^{\dagger }$, $H_{S}=\Omega b^{\dagger}b$. Here
$b^{\dagger}$ ($b$) is the bosonic creation (annihilation)
operator.

\textit{Adiabatic evolution and the Born-Oppenheimer approximation.--- }
Consider the nonresonant case where the subsystem characteristic
frequencies are {\it high} relative to the  frequencies of the
reservoirs \cite{comment}.
This implies a timescale separation as the subsystem dynamics is fast,
while the bath motion is slow. A BO type
approximation can thus be employed following two consecutive steps:
First, the fast variable is considered:
We solve the subsystem eigenproblem  fixing the reservoirs configuration,
acquiring a set of potential energy surfaces 
which parametrically depend on $B_{\nu}$.
In the second step we assume that the baths dynamics evolves on 
the ground potential surface, and solve
the vibrational heat transfer problem, form $L$ to $R$.
Next we follow this procedure using the generic Hamiltonian (\ref{eq:H}).
Beginning with the fast contribution, we diagonalize
\bea
H_{g}=H_{S}+SB; \,\,\,\, (B=B_{L}+B_{R}),
\label{eq:Hg}
\eea
and acquire the potential surface $W$. For example, for a TLS subsystem
 we resolve $\left\vert g(B_{v})\right\rangle =c_{0}\left\vert
0\right\rangle +c_{1}\left\vert 1\right\rangle$ as the ground state
of $H_{g}$; $c_{0,1}$ are the superposition coefficients, functions
of $B$ and $\epsilon$, with the eigenenergy $W=-\sqrt{(\epsilon
/2)^{2}+B^{2}} =-\epsilon/2 -B^{2}/\epsilon +B^{4}/\epsilon ^{3}+{\cal
O}(B^{6}/\epsilon^5)$.
For an  harmonic oscillator model we exactly obtain $W=\frac{\Omega }{2}-\frac{B^{2}}{\Omega}$,
including the zero-point motion.
We assume next that the total density matrix is initially factorized,
\bea
\rho(0)=\left\vert g(B_{\nu})\right\rangle \rho _{B}(0)\left\langle
g(B_{\nu})\right\vert,
\eea
where $\rho_B(0)=\rho_L \times \rho_R$;
$\rho_{\nu}=e^{-\frac{H_{\nu}}{T_{\nu}}}/{\rm
Tr_{\nu}}\big[e^{-\frac{H_{\nu}}{T_{\nu}}}\big]$ is the
equilibrium-canonical distribution function of the $\nu$ bath. 
Time evolution is dictated by the Liouville equation,
\bea
\rho (t) &=&e^{-iHt}\rho (0)e^{iHt} \nonumber \\
&\approx &\left\vert g(B_{\nu})\right\rangle e^{-iH_{BO}t}\rho
_{B}(0)e^{iH_{BO}t}\left\langle g(B_{\nu})\right\vert,
\eea
where the second step is justified under the BO
approximation with the effective Hamiltonian
\bea
H_{BO}=H_{L}+H_{R}+W.
\label{eq:HBO}
\eea
Thus, the \emph{reduced density matrix} of the
reservoirs, $\rho _{B}(t)=$Tr$_{S}\rho (t)$, where the trace is performed
over the subsystem degrees of freedom, evolves as
\bea
\rho _{B}(t)=e^{-iH_{BO}t}\rho _{B}(0)e^{iH_{BO}t}.
\label{eq:rho}
\eea
In the present scheme we thus propagate the {\it bath} coordinates
along the {\it subsystem} potential energy surface $W$,
and an explicit study of the subsystem motion is not required,
unlike the typical situation in other approaches \cite{Lepri,Dhar,MasterD}.
We identify the operator $W$ as an interaction term
directly connecting the two reservoirs. Note that in the
original model, Eq. (\ref{eq:H}),
$V$ is linear in $B$, additive in the $L$ and
$R$ coordinates. In contrast, under the BO
approximation we obtain a
potential energy surface $W$ which is often nonlinear in $B$,
mixing the left and right reservoirs' coordinates in a nontrivial way.

\textit{Heat current.--- }
The heat current operator, between the two reservoirs, can be defined as \cite{Wucurr}
\bea
\hat J_{L}=i[H_{L},W]/2.
\label{eq:Jop}
\eea
For example, for a harmonic subsystem we recover
$\hat J_{L}=-\frac{1}{2\Omega }(BP_{L}+P_{L}B)$, 
while for a TLS subsystem  $\hat J_{L}\approx - \frac{1}{2\epsilon }(BP_{L}+P_{L}B)$
when $B/\epsilon \ll 1$; $P_{L}=i[H_{L},B_{L}]$. The current
operator in both cases is identical in the first order of $B/\epsilon$.
Generally, the expectation value of the current is
\bea
J_{L}(t) =\text{Tr}[\hat J_{L}\rho _{B}(t)]
=\text{Tr}[e^{iH_{BO}t}\hat J_{L}e^{-iH_{BO}t}\rho _{B}(0)],
\label{eq:Jav}
\eea
where the left expression is written in the Sch\"oredinger picture; the
second is in the Heisenberg representation. 
The trace is performed over the two baths degrees of freedom.

\textit{First Order Current---} When system-baths couplings, absorbed in $W$, are weak, the
time evolution operator can be approximated by the first order term
\bea
e^{-iH_{BO}t}=e^{-i(H_{L}+H_{R})t}\left(1-i\int_{0}^{t}W(\tau )d\tau \right),
\label{eq:weak}
\eea
and the current (\ref{eq:Jav}) reduces to 
\bea
J_{L}(t)=-i\int_{0}^{t}\text{Tr}\{[\hat J_{L}(\tau ),W]\rho _{B}(0)\}d\tau,
\label{eq:curr}
\eea
where $W(\tau)$ and $\hat J_{L}(\tau)$ are interaction picture
operators, $A(t)=e^{iH_Bt}A e^{-iH_Bt}$ with $H_B=H_L+H_R$. We are
mostly interested in  steady state quantities,
$J=J_{L}(t\rightarrow \infty )$, if the limit exists. 
This expression can be further customized
by using a diagonal form for the reservoirs,
e.g., for the $L$ bath we write, $H_{L}=\sum E_{k}\left\vert k\right\rangle
\left\langle k\right\vert$, and by expanding the potential surface in the left bath ($L$) and 
right ($R$) bath  operators, functions of $B_{\nu}$,
\bea W=\sum_{a,b}L^{a}\otimes R^{b}=\sum_{a,b}\sum_{k,m}
\sum_{p,s}L_{k,m}^{a}R_{p,s}^{b}\left\vert kp\right\rangle
\left\langle m s\right\vert.
\label{eq:W}
\eea
$|k\rangle$ and $|m\rangle$ are the many body states
of the left reservoir with energies $E_k$ and $E_m$;
$|p\rangle$ and $|s\rangle$ are the many body states
of the right reservoir with energies $E_p$ and $E_s$.
The interaction $W$ sums (nonseparable) contributions from the two reservoirs,
$a$ and $b$ are integers. For example,
for the harmonic subsystem with bilinear coupling 
$W =-B^2/\Omega=-(B_L^2+B_R^2 +2B_LB_R)/\Omega$.
It can be shown that terms containing
either $L$ or $R$ operators do not add to the current, as only mixed terms account.
Therefore, in the case of a harmonic subsystem a single term contributes
to (\ref{eq:W}) with $L^1=iB_L\sqrt{2/\Omega}$ and $R^1=iB_R\sqrt{2/\Omega}$. 
Back to (\ref{eq:curr}),
employing (\ref{eq:W}), we  accede to the steady state heat current
\bea J &=&\frac{2\pi }{Z_{L}Z_{R}}\sum_{a,b}\sum_{k,m,p,s}E
_{_{k m}}\left\vert L_{km}^{a}\right\vert ^{2}\left\vert
R_{ps}^{b}\right\vert ^{2} \nonumber\\
 &&\times \delta (E _{km}+E
_{ps})e^{-\beta _{L}E_{k}-\beta _{R}E_{p}} \nonumber\\
 &=&\frac{2\pi
}{Z_{L}Z_{R}}\sum_{a,b} \sum_{k,m,p,s}
 \big ( \left\vert L_{km}^{+a}\right\vert ^{2}\left\vert
R_{ps}^{-b}\right\vert ^{2}-\left\vert L_{km}^{-a}\right\vert
^{2}\left\vert R_{ps}^{+b}\right\vert ^{2} \big ) \nonumber\\ 
&&\times  \left\vert E_{km}\right\vert
\delta (\left\vert E_{km}\right\vert -\left\vert E
_{ps}\right\vert )e^{-\beta _{L}E_{k}-\beta _{R}E_{p}},
\label{eq:curr2} \eea
where $E_{km}=E_{k}-E_{m}$ and, e.g., $Z_{L}=\sum_k e^{-\beta
_{L}E_{k}}$ is the $L$ bath partition function.  
Here $L_{km}^{+a}$ ($L_{km}^{-a}$) denotes  matrix
elements when $E_{k}>E_{m}$ ($E_{k}<E_{m}$). We identify next the
Fermi-like golden rule excitation ($+$) and relaxation ($-$) rates, e.g., at the $L$ contact, by
\bea
k_{L}^{\pm a}(\omega)=2\pi \sum_{k,m}|L_{k,m}^{+a}|^2\delta(E_k-E_m\mp\omega)\frac{e^{-\beta_LE_k}}{Z_L},
\label{eq:rate}
\eea
satisfying  detailed balance, $k_{L}^{+a}(\omega)=k_{L}^{-a}(\omega)e^{-\beta_L \omega}$.
We can therefore reduce Eq. (\ref{eq:curr2}) into the compact form
\bea
J=\frac{1}{2\pi} \sum_{a,b}\int_0^{\infty} \omega d\omega
 k_L^{-a}(\omega) k_R^{-b}(\omega) ( e^{-\beta_L\omega}-e^{-\beta_R\omega})
\label{eq:currF}
\eea
where the sum over $a$ and $b$ is determined given a particular $W$.
This is the main result of our paper.
We refer to this expression as the "generalized Landauer formula" \cite{Land},
as the net heat current is given by the difference between left-moving 
and right-moving excitations.
Nevertheless, our formula can incorporate anharmonic interactions,
absorbed in the rates $k^{\pm a}_{\nu}(\omega)$ unlike the original treatment  \cite{Land}.
We emphasize the broad status of Eq. (\ref{eq:currF}).
It was derived without specifying the
subsystem Hamiltonian or the system-bath interaction form, both contained in $W$. It is valid
as long as (i) there exists a timescale separation between the subsystem motion (fast)
and the reservoirs dynamics (slow), and (ii) system-bath interaction is weak,
see Eq. (\ref{eq:weak}).
In what follows we apply Eq. (\ref{eq:currF}) on some models of particular interest:
a fully harmonic model, a nonlinear model with  strong system-bath interactions, and 
utilizing a spin subsystem, representing a nonlinear impurity.


\textit{Harmonic model.---}
We consider first a harmonic model, $H=H_L+H_R+V_L+V_R+H_S$, with
\bea
&&H_{\nu}=\sum_{j\in \nu} \omega_{j}b_{\nu,j}^{\dagger}b_{\nu,j}; \,\,\,\,\,\,\,
H_S=\Omega b^{\dagger}b, \,\,\,\,\ V_{\nu}=S B_{\nu},
\nonumber\\
&& S=(b^{\dagger}+b), \,\,\,\,
B_{\nu}=\sum_{j\in \nu} \lambda_{\nu,j}(b_{\nu,j} + b_{\nu,j}^{\dagger}),
\label{eq:HHar}
\eea
and show that Eq. (\ref{eq:currF}) reduces to the standard elastic limit 
 \cite{Land,Segalcond}. 
Here the subsystem comprises a single mode of frequency $\Omega$.
$b_{\nu,j}^{\dagger}$ ($b_{\nu,j}$) are the creation
(annihilation) operators of the mode $j$ in the $\nu$ bath,
$b^{\dagger}$ and $b$ are the respective subsystem operators.
$\lambda_{\nu,j}$ are system-bath interaction energies,
$S$ is a subsystem operator.
The expectation value of the current can be calculated either by
following Eq. (\ref{eq:curr}) in the long time limit, or by
directly applying Eq. (\ref{eq:currF}), as we do next.
In the occupation number representation the many body states of the $L$ reservoir are
$|m\rangle=| m_1, m_2... m_l... m_N \rangle$ with
$m_l$ excitations for the $l$ mode.
Since  $W=-B^2/\Omega$, the relevant matrix elements in (\ref{eq:W}) are
%
$|L_{km}^1|=\sqrt{\frac{2}{\Omega}}\sum_l \lambda _{L,l } \left(\sqrt{m_l+1}\delta
_{k_l,m_l+1} +\sqrt{m_l}\delta_{k_l,m_l-1} \right)$. 
%
An analogous expression exists for $R_{ps}^1$.
We thus identify the excitation and relaxation rates in Eq. (\ref{eq:rate}) by 
$k_{\nu}^{+1}(\omega)=\frac{2}{\Omega}\Gamma_{\nu}(\omega)n_{\nu}(\omega)$
and $k_{\nu}^{-1}(\omega)=\frac{2}{\Omega}\Gamma_{\nu}(\omega)[n_{\nu}(\omega)+1]$,
respectively, where $n_{\nu}(\omega)=\left[ e^{\omega/T_{\nu}}-1\right]^{-1}$ is the Bose-Einstein distribution
function and
$\Gamma_{\nu}(\omega)=2\pi \sum_{j\in \nu}\lambda_{\nu,j}^2 \delta(\omega-\omega_j)$.
Using these rates the current (\ref{eq:currF})  reduces to
\bea
J=\frac{2}{\pi}\int_0^{\infty} \frac{\Gamma_L(\omega) \Gamma_R(\omega)}{\Omega^2}
\left[n_L(\omega) - n_R(\omega) \right] \omega d\omega.
\label{eq:LandBO}
\eea
This is the Landauer's formula for heat conduction  \cite{Land} in the BO limit; assuming
the system frequency is above the baths spectral window, further
utilizing the weak-coupling approximation [Eq. (\ref{eq:weak}].
Beyond this limit, the heat current for the harmonic model
(\ref{eq:HHar}) is exactly given by
\bea
J=\frac{2}{\pi}\int  {\cal T}(\omega)[n_L(\omega)-n_R(\omega)] \omega d\omega,
\label{eq:Jh}
\eea
with the transmission coefficient 
${\cal
T}(\omega)=\frac{\omega^2 \Gamma_L
\Gamma_R}{[(\omega^2-\Omega^2)^2+(\Gamma_L+\Gamma_R)^2\omega^2]}$ \cite{Segalcond}.
The rate $\Gamma_{\nu}$ has been defined above Eq. (\ref{eq:LandBO}); 
for convenience we discard the direct reference to frequency.
In the weak coupling limit, $\Gamma_{\nu}<\Omega$, the transmission
coefficient is sharply peaked around $\Omega$. In the nonresonant case $\Omega\gg \omega_c$,
where $\omega_c$ is the reservoirs cutoff frequency, 
${\cal T}(\omega)\sim \frac{\Gamma_L(\omega) \Gamma_R(\omega)}{\Omega^2}$
and Eq. (\ref{eq:LandBO}) is recovered.
In the opposite limit, when the baths spectral window overlap with
the molecular vibrations, $\omega_c\gg \Omega$, Eq. (\ref{eq:Jh}) reduces into a resonant energy transfer 
expression,
%
$J= \Omega\frac{\Gamma_L\Gamma_R}{\Gamma_L+\Gamma_R}[n_L(\Omega)-n_R(\Omega)]$.
%
Here $\Gamma_{\nu}$ is calculated at the (local oscillator) frequency $\Omega$.
This expression describes a  hopping motion, 
with energy flowing sequentially from the $L$  bath into the
subsystem, then into the $R$ contact. This  process
is dictated by the subsystem energetic window,  yielding $J\propto \Omega$.
In contrast, Eq. (\ref{eq:LandBO}) accounts for a coherent, 
deep tunneling energy transfer mechanism, and the current decays 
with the energetic barrier, $J\propto 1/\Omega^2$.

\textit{Anharmonic models.---}
We generalize next the harmonic result by modifying the
model (\ref{eq:HHar}), adopting an exponentially repulsive interaction,
\bea
B_{\nu}= e^{-\sum_{j}\lambda_{\nu}(b^{\dagger}_{\nu, j}+ b_{\nu,j})},
\eea
appropriate for the relevant nonresonant case \cite{NitzanBook}.
As before, diagonalizing $H_g=H_S+ V_{L}+V_{R}$ we obtain the potential surface $W=-B^2/\Omega$; $B=B_{L}+B_R$,
controlled by the bipartite term $2B_LB_R/\Omega$.
For simplicity, we assume an (identical) Einstein-type model for the reservoirs spectra, 
represented by a single frequency $\omega_B$. Under this assumption
the relevant excitation/relaxation rates [Eq. (\ref{eq:rate})] are given by
\bea
k_{\nu}^{\pm 1}(\omega)&=&\frac{2}{\Omega}\sum_{l=0}^{\infty} 2\pi\frac{\lambda_{\nu}^{2l}}{ l!}\sum_{s=0}^l 
\frac{l!}{(l-s)!s!}[n_{\nu}(\omega_B)+1]^s 
\nonumber\\
&\times&
n_{\nu}(\omega_B)^{l-s}\delta(\mp \omega -(2s-l)\omega_B).
\eea
%
%
Assuming $\lambda$ is small, we enclose only single-phonon and two-phonon contributions
in (\ref{eq:currF}), yielding the heat current
\bea
&&J=\frac{8\pi}{\Omega^2}\Big\{ \omega_B\lambda_L^2\lambda_R^2 (n_L-n_R) 
\nonumber\\
&&+\frac{(2\omega_B)\lambda_L^4\lambda_R^4}{4} \left[ n_L^2(n_R+1)^2-n_R^2(n_L+1)^2\right] 
\Big\}.
\eea
The Bose-Einstein functions are evaluated at the frequency $\omega_B$.
This expression presents a generalization to the harmonic result (\ref{eq:LandBO}),
accommodating multiphonon  processes; the second term describes tunneling of a two-phonon combination.

\begin{figure}
\hspace{2mm}
{\hbox{\epsfxsize=70mm \epsffile{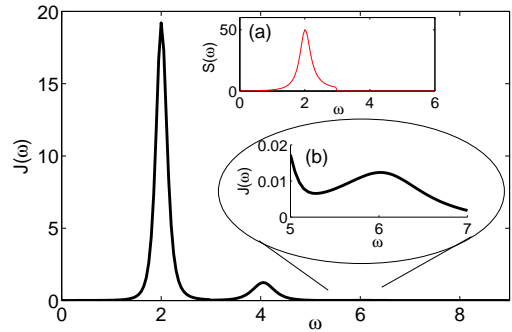}}}
\caption{Frequency components of the heat current $J(\omega)$,
for a spin subsystem bilinearly coupled to heat baths.
Multiphonon processes are observed, see also panel (b).
The parameters $\epsilon=6$, $T_L=1$, $T_R=0.5$ were used with
the bath spectral function $S(\omega)$, depicted in panel (b), identical at the two ends.
} \label{FigJ}
\end{figure}

The starting point of our next anharmonic model is again Eq.
(\ref{eq:HHar}), utilizing  a two-level subsystem, $H_S=\frac{\epsilon}{2}\sigma_z$, 
$S=\sigma_x$, representing a nonlinear impurity bilinearly coupled to bath phonons.
In this case we resolve $W=-\sqrt{B^2+\epsilon^/4}$, thus
the first order current is identical to the harmonic result (\ref{eq:LandBO}),
with $\Omega$ replaced by $\epsilon$. 
Incorporating the next term in the expansion, $W\sim \epsilon/2 -B^2/\epsilon+B^4/\epsilon^3$, we get
\bea
&& J=\frac{1}{2\pi} \int_0^{\infty} \omega d\omega (e^{-\beta_L \omega}-e^{-\beta_{R}\omega})
\big[ k_L^{-1}(\omega)k_R^{-1}(\omega) +
\nonumber\\   
 && k_L^{-2}(\omega)k_R^{-2}(\omega) +
 k_L^{-3}(\omega)k_R^{-1}(\omega)  + k_L^{-1}(\omega)k_R^{-3}(\omega)  \big].
\label{eq:Jspin}
\eea
The first term in the square brackets describes a single phonon (harmonic) process, 
proportional to $1/\epsilon^2$. The other terms  collect contributions 
from multiphonon processes. For example, the second element accounts for 
the absorption of two phonons in the left bath, followed by an
emission of these phonons at the other end, with, e.g.,
\bea
&&k_L^{-2}(\omega)\propto \frac{1}{\epsilon^3}\sum_{l,l'}(1+n_l)(1+n_{l'})\delta(\omega-\omega_l-\omega_{l'})
\nonumber\\
&& + \frac{1}{\epsilon^3}\sum_{l,l'}2(1+n_l)n_{l'}\delta(\omega-\omega_l+\omega_{l'}).
\eea
The last two contributions in (\ref{eq:Jspin}) convene three-phonon processes, where, e.g., a single mode 
from the left bath decays into three excitations at the right side.
Fig. \ref{FigJ} presents the frequency components of the heat current 
(the integrand of Eq.  (\ref{eq:Jspin})), where for simplicity we assume a 
spectral density $\mathcal {S}(\omega)=\sum_j \lambda_{\nu,j}^2\delta(w-\omega_j)$ peaked around a specific
bath frequency $\omega_B=2$  with a hard cutoff at $\omega=3$, see panel (a).
We identify three contributions to the current: A dominant, single-phonon element
at $\omega\sim \omega_B$, a weaker two-phonon contribution,
and a rudimentary three-phonon current,  see panel (b).
In the presence of a spatial asymmetry these high order terms are responsible for the thermal rectification effect \cite{Rectif}. 
Note that in the resonant regime, when a hopping mechanism dominates,
the heat current across harmonic junctions is {\it higher} than its anharmonic
counterpart \cite{MasterD} due to a saturation effect. 
In contrast, in the nonresonant case anharmonicity {\it enhances} the thermal 
current  due to the participation of multiphonon processes.
Similar observations were obtained in a study of classical heat flow in molecular junctions \cite{Yun}.


\textit{Summary.---}
We have presented here a generally applicable Born-Oppenheimer type formalism for describing  
thermal energy transfer in the off-resonant case,
where an impurity object has a characteristic frequency above the (populated) bath modes.
In this limit energy propagates across the structure in a tunneling-like motion, keeping the
subsystem population intact.
In the weak coupling limit we derived a compact expression for the thermal current, 
bearing the structure of a generalized Landauer relation, yet incorporating multiphonon effects.
In the harmonic limit our formula reduces to known results. We have also applied it
onto nonlinear models: Incorporating molecular anharmonicity or assuming 
short range interactions we reach
simple analytic expressions for the heat current, 
reflecting the underling transport mechanism.
The new method described here is complementary to kinetic approaches
that are typically valid in the resonant case.
By incorporating quantum effects and nonlinearity it opens new
ways for describing vibrational or electronic energy 
dynamics in organic materials \cite{Scholes}, biomolecules \cite{Leitnerbook} and superconductors 
\cite{Pekola}.

L. -A. Wu has been supported by the Ikerbasque Foundation Start-up, the
CQIQC grant and the Spanish MEC (Project No. FIS2009-12773-C02-02).
DS was supported by NSERC.



\begin{thebibliography}{9}
\bibitem{comment}
Our scheme is valid for Ohmic baths, if their temperatures are low, 
below he subsystem energy spacing.
In such cases bath modes overlapping with subsystem  vibrations
are not populated and  the dynamics is controlled
by off resonance bath modes.
\end{thebibliography}
\end{document}